\documentclass[12pt]{article}

\input epsf
\def\beq{\begin{eqnarray}}
\def\eeq{\end{eqnarray}}

\def\k{{\bf k}}

\def\lsim{\mathrel{\rlap{\lower3pt\hbox{\hskip0pt$\sim$}}
     \raise1pt\hbox{$<$}}}         
\def\gsim{\mathrel{\rlap{\lower4pt\hbox{\hskip1pt$\sim$}}
     \raise1pt\hbox{$>$}}}         

\usepackage{amsmath}
\usepackage{amsfonts}

\begin{document}

\begin{titlepage}

\thispagestyle{empty}

\begin{flushright}
{NYU-TH-08/10/62}
\end{flushright}
\vskip 0.9cm

\centerline{\Large \bf Effective Lagrangian and Quantum Screening  }
\vskip 0.2cm
\centerline{\Large \bf in   Charged  Condensate}                    

\vskip 0.7cm
\centerline{\large Gregory Gabadadze and Rachel A. Rosen}
\vskip 0.3cm
\centerline{\em Center for Cosmology and Particle Physics,  
Department of Physics,}
\centerline{\em New York University, New York, 
NY  10003, USA}

\vskip 1.9cm

\begin{abstract}

A condensate of charged scalars in a neutralizing background 
of fermions (e.g., condensed helium-4 nuclei in an  electron background  
in white  dwarf cores) is investigated further. We discuss an  
effective Lagrangian  approach to this system  and  
show that the strong screening of  an  electric charge found 
previously in arXiv:0806.3692 in a  mean-field approximation,  
is a consequence of a cancellation due to a phonon.  
The resulting propagators contain terms that strongly 
modify their infrared behavior. Furthermore, we evaluate a one-loop 
fermion quantum  correction to the  screened  potential, and find that it is 
also  suppressed  by the phonon  subtraction.  Therefore, charged 
impurities  (e.g., hydrogen or helium-3 nuclei)  will be  
screened  efficiently by the condensate.

\end{abstract}


\end{titlepage}

\newpage

\begin{center}
{\large \bf 1. Introduction and Summary}
\end{center}

\vskip 0.3cm

Consider a system of negatively charged fermions (e.g., electrons) 
of mass $m_f$ and of number-density $J_0$,  and positively  
charged  scalars (e.g., helium-4 nuclei) of mass $m_H$, 
with  the net charge of the system being zero.  At densities 
$J_0\simeq (0.5 - 5\,{\rm MeV})^3$ the 
average inter-fermion separation, $d\sim J_0^{-1/3}\sim 
(400-40\,fm)$,  is  much smaller than  the atomic scale $\sim 10^5\,fm$,  
while being   much greater than  the nuclear scale $\sim 1\,fm$.  
Therefore, neither atomic nor nuclear effects would matter 
at such  densities.  The fermions would form a degenerate 
gas with the Fermi  energy exceeding inter-fermion repulsion energy, 
which can be ignored. These are the conditions thought to be 
present in cores of helium white dwarfs. 

As a  helium dwarf star cooled  from $\sim 10^6-10^7 \,K$, down  
to lower temperatures, the helium-4 nuclei could have 
formed a  crystalline  structure. However, the zero-point 
oscillations of helium-4 ions at these densities    
would  ``melt''  such a  crystal, and it was 
proposed \cite {GGRR1,GGRR3},  that the helium-4 ions could 
instead condense into a macroscopic state of large 
occupation number -- the charged condensate --  specific 
properties of which were discussed in detail in 
\cite {GGRR3}.

In particular, a static potential between  probe 
charges placed in the condensate were calculated in 
\cite {GGRR3} in the mean-field approximation,  and were  
shown to be suppressed  by the exponential factor, 
$\sim {\rm exp} (-Mr){\rm cos} (Mr)/r$, 
where the scale $M$  is defined  in terms of 
the electric charge $e$,  mass $m_H$, and 
number-density $J_0$: 
\beq
M\equiv  (2 e^2 m_H J_0)^{1/4}\,> J_0^{1/3}\,.
\label{M}
\eeq
Although this strong  screening may well be a reason why the 
condensation of charged bosons takes place in the first place, 
the following issues emerge in this regard:

(i) It may seem that the exponent ${\rm exp} (-Mr)$ is due to a 
state  of mass $M$.  The distance scale $1/M$ is shorter  than  
the  average inter-particle separation -- an effective short-distance 
cutoff of the low-energy theory.  Then, a state of mass $M$, if existed,  
would have been beyond the scope of   the low-energy  
field theory description, and the above-cited potential 
would have been unreliable.

(ii) The momentum-space static potential, the 
Fourier transform of which is proportional to 
$\sim {\rm exp} (-Mr){\rm cos} (Mr)/r$, 
reads as follows (see Section 3)
\beq
G(\omega=0, \k) \simeq \left ( \k^2 + m_0^2 +{4M^4\over \k^2} \right 
)^{-1}\,,
\label{staticI}
\eeq
where, $m_0^2$  stands for a certain quantity such that 
$|m_0^2|\ll M^2$ (see Ref. \cite {GGRR3} and Sections 3,4 below). 
The term $M^4/k^2$  gives rise to a significant modification of the 
Green's function in the infrared.  The interpretation 
of such an infrared-sensitive term in the denominator of a 
propagator calls for an explanation.

(iii) In the $m_H \to \infty $ limit  one would expect 
the heavy scalars (helium-4 nuclei) to decouple.
It is not exactly clear from (\ref {staticI}) how
such a decoupling takes place, and what is its interpretation.

\vspace{0.1cm}

The above three issues, can be clarified,    
as we will summarize below, and show  in this paper.
The points  (i) and (ii)  get resolved by observing that   
the Green's function (\ref {staticI}) can formally 
be decomposed as follows:
\beq
G(\k, \omega=0) \simeq   {1\over \k^2+m_0^2}  -  {1\over  
\k^2 + m_0^2 + {\k^2(\k^2+m_0^2)^2/4M^4}}\,.
\label{photonphononI}
\eeq
The first term on the r.h.s. can be thought of as  
an instantaneous screened Coulomb 
potential, while the second one can be interpreted as 
a potential due to a phonon.  The phonon in this case is a  
collective excitation  of motion of 
charged  scalars within the fermion background.  As was shown in 
\cite {GGRR3}, the phonon is a light excitation  that  belongs 
to the spectrum  of a low-energy effective field theory.
It's just the cancellation due to this light mode  
that gives rise to the exponential  ${\rm exp} (-Mr)$, 
and not a hypothetical state of mass $M$. 

Before we turn to the point (iii) let us make two important 
comments.  First, we should note that the second term in  
(\ref {photonphononI})  has three poles. 
The residue of the pole at $\k^2+m_0^2=0$  cancels exactly with 
that from the first term of  (\ref {photonphononI}).  
The remaining two poles, describe 
both the heavy state of mass $\sim m_H$, 
and a light state, which  actually is 
the phonon \cite {GGRR3}. For simplicity of the discussions, 
in  this  work we'll be using a somewhat arbitrary 
language  by calling  the whole second term in  (\ref {photonphononI}) 
the phonon contribution.

Second, we note that the form of  (\ref {photonphononI}) may be 
puzzling itself: the negative sign in front of the second 
term on the r.h.s. may  seem 
to be suggesting  that  the phonon is a ghost. This indeed would 
have been a ghost contribution   
if we were to obtain such a term in a spectrum of a Lorentz-invariant theory.
Here, the background solution fixes the Lorentz frame, and the spectrum 
of small perturbations is not Lorentz-invariant \cite {GGRR1}.  
Because of this, the negative sign in front  of the phonon Green's 
functions is not a signature of a ghost. In particular, as we'll show 
in Section 3, the Hamiltonian for the  fluctuations  that give 
rise to (\ref {photonphononI}) is positive semi-definite!  
Moreover, the measure of 
Lorentz-breaking is the scale $M$. In the $M\to 0$ limit,
the phonon term in (\ref {photonphononI}) vanishes, as one would expect.

\vspace{0.1cm}

Let us now turn to the point  (iii).  In the limit $m_H \to \infty$, 
which implies $M \to \infty$,  the phonon effects should  go away.
This is  reflected in (\ref {photonphononI}) as vanishing of the 
whole potential, which is just 
a consequence of using the static approximation, 
and can be understood in  the following  way:
The phonon mixes with the timelike component of the gauge field, and 
due to this acquires an instantaneous part. Then, the instantaneous parts 
in (\ref {photonphononI}) cancel between the gauge and photon contributions.
However, the dynamical part of the phonon is also reducing to 
zero, because the group velocity of the phonon vanishes in 
the $m_H \to \infty$ limit,  as  we will show in Section 3.

\vspace{0.1cm}

Having the above properties clarified, 
in this work we will study further the static charge screening 
in charged condensate.  The physical question  is that of  static  
interactions  between charged impurities placed into the condensate.
As we mentioned above, the scalar condensate gives rise to the 
suppression of the static potential  by a factor  ${\rm exp} (-Mr)$. 
Inclusion of the fermion fluctuations via the mean-field 
Thomas-Fermi  method,  does not change the above 
result significantly \cite {GGRR3}.

In Section 4  we go beyond the mean-field approximation and include
a one-loop quantum correction due to the polarization diagram.  
This diagram is suppressed by an additional power of the 
electromagnetic coupling constant $\alpha_{\rm em}=e^2/4\pi$, 
and one would expect the quantum correction to be insignificant. 

However, this is not the case for the following subtle reason. 
The one-loop correction
introduces branch cuts in the Green's function (\ref {photonphononI}),
which give rise to additional contributions to the static potential  
in the  position space.  These 
additional terms have oscillatory nature  with a {\it power-like}  
decaying envelope. Even though they  are formally suppressed by 
$ {\cal O}(\alpha^2_{\rm em})$,  to  a good approximation they end 
up being  $ {\cal O}(1)$,  and can dominate over the exponentially  
suppressed term at sufficiently large distances.

Such an oscillatory potential is know in non-relativistic 
fermion systems as the Friedel potential (for original 
references see \cite {Walecka}), which was generalized to relativistic 
fermions by Sivak \cite {Sivak} and by Kapusta and Toimela \cite {KapustaT}. 
Here we perform  analogous calculations for the 
charged condensate and find that the phonon subtraction 
suppresses  further the Friedel-like
potential by the factor $J_0^{8/3}/M^8\sim 10^{-23}$. 
In spite of this,  for reasonable separations between the test 
particles,  the  Friedel-like  potential  dominates 
over the  exponential one. 

Therefore, charged impurities, such as  hydrogen or helium-3 nuclei, 
once placed within the charged condensate,  will interact 
very weakly with the ambient condensate and with each other.
The impurities are screened very efficiently  by the condensate!

\vspace{0.1cm}

Let us make a few comments on the literature. Condensation of 
non-relativistic charged scalars has a long history, 
the original works being those by  Schafroth \cite {Shafroth}, 
in the context of superconductivity, and by 
Foldy \cite {Foldy},  in a more general setup. 
An almost-ideal Bose gas approximations was 
assumed in those studies. For this assumption to be true, 
densities had to be taken high-enough  to make the  average inter-particle  
separation  shorter than  the boson Bohr radius \cite {Foldy}. 
In terms of our parameters this would be the case 
if $J_0^{1/3}\gsim  \alpha_{\rm em} m_H$. 

However, for the helium-electron system the above condition  
would translate into super-high densities,   at which 
nuclear interactions  become significant. 
Instead, in this work we're studying charged condensation in the opposite 
regime,  $J_0^{1/3}\ll \alpha_{\rm em} m_H$, where  the 
nuclear forces play no role.  Moreover,  our mechanism of screening the 
electric charge is different, as it is based on  cancellation by 
the phonon.  

A possibility of having a changed condensate 
in helium white dwarfs was previously discussed in Ref.
\cite {71}\footnote{We thank Andrei Gruzinov for 
bringing this  paper  to our attention. Regretfully, this happened  
after our work  on charge condensation in 
white dwarfs \cite {GGRR3} was already published. We use this 
opportunity here to emphasize differences between our 
approach and that of  \cite {71}.}, where the 
condensation  was studied  using an approximate variational 
quantum-mechanical calculation in conjunction  with numerical 
insights  in a strongly-coupled regime of 
electromagnetic interactions. A degree of reliability of such a scheme 
is hard to assess\footnote{Furthermore, using the ordinary neutral 
Bose-Einstein (BE) condensation  
to describe the charged condensate, as it is done in a number of 
works in  the literature,  is unjustified.  Properties of a 
neutral BE condensate  differ significantly from those of   
the charged condensate considered  here.  For instance, 
specific heat at moderate temperatures  in the  former is due 
to a phonon gas, while in the latter it is due to the 
degenerate electrons (see, e.g., \cite {GGRR3}).}. Finally, 
in the context of a relativistic field theory  condensation of scalars 
was discussed in, e.g., Refs. \cite {Linde,Kapusta,Haber}, 
some of the results of which we cite below.

A. Dolgov, A. Lepidi, and G. Piccinelli \cite {Dolgov}
(appeared on archive the same day with this paper)  calculated a one-loop 
correction  to the photon  self-energy in the background  of charged  
condensate at nonzero  temperature. These author find similar infrared 
modifications of the static potential as we did. Moreover, their results  
also include  nonzero temperature effects, and in that part,
are more general  than ours.  On the other hand, our emphasis 
is on understanding of these results in terms of the phonon 
and  effective Lagrangian language.

\vspace{0.3cm}
\begin{center}
{\large \bf 2. Effective Lagrangian and its Solution}
\end{center}
\vspace{0.3cm}

In this Section we discuss an effective  Lagrangian 
description of charged condensation. We will treat the 
bosonic and fermionic degrees of freedom separately. This has two reasons: 
(a) There is a significant  disparity of physical 
scales in these two sectors; 
(b) Due to the Fermi surface, the dynamics for fermionic fluctuations 
should be though of in terms of near-the-Fermi-surface degrees of freedom. 
For instance, when the  energy of these excitations scales to zero, their 
momenta scale to the Fermi surface \cite {Polchinski}. To be able to describe 
the dynamics of the near-the-Fermi-surface  fluctuations, 
including  their one-loop quantum polarization effects, 
we will adopt for them a  microscopic Lagrangian with a relativistic 
chemical potential
\beq
{\bar \psi}(i\gamma^\mu D_\mu -m_f)\psi + \mu_{f} \psi^+ \psi\,, 
\label{lagrF}
\eeq
where the covariant derivative for fermions (electrons) 
is defined  as $D_\mu = \partial_\mu +ie A_\mu $,
and $\mu_{f}$ denotes the fermion chemical potential, 
that equals to the Fermi energy at zero temperature 
$\mu_{f}=\epsilon_F= [(3 \pi^2 J_0)^{2/3}+m_f^2]^{1/2}$ .

For the bosonic sector, on the other hand,  we will employ the 
effective order-parameter description. In particular we will look
for classical solutions of the equations of motion of the effective 
order-parameter Lagrangian.

How could a classical solution 
describe the condensation which is 
an inherently quantum phenomenon?  Denote the particle creation and 
annihilation  operators  by $a_0^+$ and $a_0$ respectively;    
then,  the  quantum-mechanical noncomutativity of these operators, 
$a_0^+ a_0 - a_0 a^+_0 \sim \hbar$, becomes an insignificant 
effect of order ${\cal O}(\hbar/N)$, when the number of particles 
in the condensate state,  $\langle a_0^+ a_0 \rangle  \sim N$,  
is large enough,  $N\gg 1$. Thus, the classical description 
of the coherent state with a large occupation number 
-- the condensate -- should be valid to a good accuracy \cite {LL}. 
On the other hand, collective excitations 
of the condensate itself should be quantized in a conventional manner.

The above arguments lead to the following decomposition of the 
order-parameter operator describing the condensate:
\beq
\Phi = \Phi_{cl} + \delta \Phi\,,
\label{classquant}
\eeq
where $\Phi_{cl}$ denotes just a classical solution 
of the corresponding equations of motion, 
and describes the condensate of many zero-momentum 
particles, while $\delta \Phi$ should describe their 
collective fluctuations. 

\vspace{0.1cm}

In the non-relativistic approximation in  $m_H$, 
the effective order-parameter Lagrangian that is consistent 
with the translational, rotational, Galilean
and the global $U(1)$ (scalar number) symmetries, as well as with 
local gauge invariance,  can be written  as follows:
\beq
{\cal L}_{eff} = {\cal P} \left (  
 {i\over 2} ( \Phi^*  D_0 \Phi -  (D_0 \Phi)^* \Phi)-
{| D_j  \Phi|^2  \over 2m_H} \right )\,,
\label{Leff}
\eeq
where $D_0 \equiv (\partial_0  - i2e A_0)$, 
$ D_j \equiv ( \partial_j - i 2e A_j) $, while  ${\cal P}(x)$
stands for a general polynomial function of its argument.
The coefficients of this polynomial are dimensionful numbers that 
are inversely proportional to powers of a short-distance 
cutoff of the effective  field theory, 
${\cal P}(x) = \sum^{\infty}_{n=0} c_n (x^n / \Lambda^{3n})$.

Such a Lagrangian was first proposed by Greiter, Wilczek and Witten 
(GWW) \cite {GWW} in a context of superconductivity. 
The requirements  leading  to (\ref {Leff}) are 
the following: (a) In the lowest order in fields it gives rise to 
the standard Schr\"odinger equation for the order 
parameter\footnote{In our nomenclature electrons are ``particles'' 
while helium-4 nuclei are ``antiparticles'', this explains the choice of 
the plus  sing in  front of the first term in (\ref {Leff}).}; 
(b) It respects all the  appropriate symmetries of the physical system 
at hand, and gives rise to an operator relation  between a current 
of charge $q$ and  the momentum density  $J_j = (q/m_H) T_{0j}$.  
(c) It gives an appropriate  spectrum of Nambu-Goldstone 
bosons in the decoupling limit\footnote{One could add to the argument of 
the function ${\cal P}(x)$ in (\ref {Leff}) terms
$\mu_{NR} \Phi^*\Phi$,  $\lambda (\Phi^*\Phi)^2$, etc. The former 
could be absorbed into a gauge potential $A_0$ by a constant shift, 
but in any case, in the condensation phase, which is a point of 
the  primary  interest here, the non-relativistic chemical potential  
$\mu_{NR}$,  should be zero
(or equivalently, the relativistic chemical potential $\mu_s = 
\mu_{NR}+m_H$ should exactly equal to the particle mass $\mu_s=m_H$). 
Furthermore, the existence of the quartic term for our system 
would not play an important role as long as $\lambda\lsim 1$ and 
$J_0\ll m^3_H$.}.

The GWW effective Lagrangian can also describe charged 
condensation, as we're about to show. For this we use the representation of   
the order parameter in terms of its  modulus and the phase,
(which is well-defined everywhere since the modulus will have a 
nonzero VEV):
\beq
\Phi = \Sigma\, {\rm exp}(i\Gamma)\,.
\label{Phi}
\eeq
Using this decomposition we rewrite the argument 
of the function ${\cal P}$ and  the effective Lagrangian 
(\ref {Leff}) in the following form: 
\beq
{\cal L}_{eff} = {\cal P} \left (  
2e \Sigma^2 B_0 - 
{( \nabla_j  \Sigma )^2 + (2e)^2 \Sigma^2 B_j^2  \over 2m_H} \right )\,,
\label{Leff1}
\eeq
where we introduced a gauge-invariant field 
$B_\mu \equiv A_\mu - \partial_\mu \Gamma/2e $.  Varying   
(\ref {Leff1}) w.r.t.  $\Sigma$ and $A_0$,  
and restricting ourselves only to constant fields, 
we obtain the corresponding equations:
\beq
{\cal P}^{\prime}  \Sigma  \left (4 e B_0 - {(2e)^2B_j^2\over m_H}\right ) =0,
~~~~ 2{\cal P}^{\prime}\Sigma^2 = J_0\,, 
\label{eqs}
\eeq
where in the last equation we took into account that the 
gauge field also couples to the background fermion charge density
$eJ_0$ as follows: $-e A_0 J_0$. Note,   
that for  space-time constant fields the equation of motion  
obtained by variation of (\ref {Leff1}) 
w.r.t. $\Gamma$  is satisfied identically.

There exists  a solution to these equations for which 
\beq
2\Sigma^2 = J_0\,,~~~ B_\mu=0,~~~~{\cal P}^{\prime}(0)=1\,.
\label{sol}
\eeq 
Since on the solution the argument of 
(\ref {Leff1}) is zero,  the condition 
${\cal P}^{\prime}(0)=1$ is satisfied by any polynomial 
functions  ${\cal P} (x)$ for which the first coefficient is 
normalized to one
\beq
{\cal P} (x) = x + C_2 x^2+...\,.
\label{px}
\eeq
The above solution describes a neutral system of 
negatively  charged electrons  of charge density 
$- eJ_0$,  and positively  charged scalar (helium-4 nuclei) 
condensate  of charge density $2e\Phi^+\Phi= 2e \Sigma^2 =eJ_0$.

\vspace{0.1cm}

Let us now turn to the issue of fluctuations about the 
classical solution. In Ref.  \cite {GWW} the Lagrangian 
(\ref {Leff}) was proposed for the description of low-energy 
theory of superconductivity, where a phonon has a 
linear dispersion relation  $\omega \propto |\k|$,  for small $|\k|$. 
Such a relation  was obtained  in  \cite {GWW} by  fixing 
$\Sigma(x)$ to a constant,  but allowing for 
fluctuations of the phase $\Gamma$.    

The system  considered here is different, however. 
In the limit of switched-off Coulomb interactions (decoupling limit)  
it should reduce to a collection of {\it free} particles of mass $m_H$. 
The latter  undergoes Bose-Einstein condensation at low temperatures.
The resulting condensate  exhibits no super-conductivity/fluidity 
because of the lack of interactions between particles. 
Hence, the  dispersion relation for the first possible excitation  
in this system  should be  $\omega \simeq  \k^2/2m_H$.  

Remarkably, this  dispersion relation can be obtained from the 
GWW Lagrangian (\ref {Leff}) by allowing for the $\Sigma(x)$ field 
to have small fluctuations. 
The latter mix with the fluctuations of the 
phase, and, as a result,  gives   rise to the dispersion  relation  
$\omega \simeq  \k^2/2m_H$, as we will show in the next Section.

The above arguments suggest  that taking a non-relativistic limit 
to decouple a dynamics of  a heavy mode in the  presence of a 
condensate, is a bit subtle, as is well known in solid state theory: 
For instance, even  though the ions in a simple crystal lattice 
could be very heavy and  non-relativistic,  there still could 
exists a low-energy mode, a phonon,  
with a linear dispersion relation $\omega \sim |\k|$, associated with  
a collective motion of those heavy ions.  The decoupling of heavy 
ions in the non-relativistic limit manifest itself as  vanishing of   
the group velocity of the phonon in the infinite-ion-mass limit.

A similar effect for a  many-body system of free bosons can be 
sketched as follows: Consider a single massive state with the dispersion
$({\omega \over c})^2 = \k^2 + m_H^2c^2$. To describe a many-body system 
of such particles one could replace  $\omega $ by $\omega + \mu_s c$,  
where $\mu_s$ is the  chemical potential, and put $\mu_s$ equals to $m_H$, 
in order to describe condensation.  Then,  the dispersion 
relation  becomes $({\omega \over c})^2= \k^2 - 2 \omega m_H $.
From this, one  gets a dispersion relation  
in the non-relativistic limit: $\omega= \k^2/2m_H$, with the group velocity
$ v_{\rm gr}= |\k|/m_H $  being a ${\cal O}(1/m_H)$ effect,
consistently with ones expectation and the  heavy mass decoupling.

\vspace{0.1cm}

For convenience, in what follows  we will  start our discussions  with a 
relativistic   Lorentz-invariant  Lagrangian that 
retains the heavy scalar mode.
Normally, this mode  would not be included in 
the effective Lagrangian, and we should be careful to
separate  heavy and light modes in our calculations, 
and use only the  light ones  for the discussion of physical effects.
Thus, the full  Lagrangian  containing the relativistic 
order-parameter $\phi$, gauge boson, and  fermions read as follows:
\beq
{\cal{L}} = -{1\over 4} F_{\mu\nu}^2 + 
\vert {\tilde D}_{\mu} \phi \vert^2 - m_H^2 \phi^{\ast} \phi +
{\bar \psi}(i\gamma^\mu D_\mu -m_f)\psi + \mu_{f} \psi^+ \psi\,.
\label{lagr0}
\eeq
A  classical nonzero vacuum expectation value of 
the field $\phi$ can serve as an order parameter for the 
condensation of the helium-4 nuclei, thus describing a 
state with a large occupation number. Fluctuations of the order 
parameter are expected to describe the collective modes of 
the condensate\footnote{The Lagrangian (\ref {lagr0}) 
is more restrictive than it's needed  -- 
the effective Lagrangian  (\ref {Leff}) 
allows for a more general velocities of  propagation than 
the Lorentz-invariant one.  However, this degeneracy of 
(\ref {lagr0})  will be lifted by  the  
fermion fluctuations,  see  Sections 3 and  4.}. 

The covariant derivative  for the 
scalar reads ${\tilde D}_\mu \equiv  \partial_\mu -2ie{\tilde A}_\mu \equiv
\partial_\mu -i(2eA_\mu +\mu_s \delta_{\mu0})\,,$ where   $\mu_s$
is the  relativistic chemical potential for the scalars.
The Lagrangian (\ref {lagr0}) is invariant under 
global $U_s(1)$ transformations, responsible for the 
conservation of the number of scalars. Another global  $U_f(1)$ 
guarantees the fermion  number conservation. One linear combination of these 
two  symmetries is gauged, and the corresponding conserved 
current is coupled to the photon  field\footnote{We could add the 
quartic scalar self-interaction 
term to (\ref {lagr0}), but this won't change our results 
significantly,  as long as the quartic coupling is not strong, and 
$m_H\gg J_0^{1/3}$.}.

As before, we introduce the representation  for 
the scalar field, $\phi = \tfrac{1}{\sqrt{2}} \sigma\, 
e^{i \alpha}$, and  work in 
the unitary gauge where the phase of the scalar is set to zero,
$\alpha=0$. In this gauge, the Lagrangian density reads: 
\beq
{\cal{L}}=- {1 \over 4}F_{\mu\nu}^2 + 
{1 \over 2}(\partial_{\mu}\sigma)^2+
{(2e)^2\over 2} {\tilde A}_\mu^2 \sigma^2- {1 \over 2}m_H^2 
\sigma^2 + {\bar \psi}(i\gamma^\mu D_\mu -m_f)\psi +\mu_f 
{\psi^+} \psi \,.
\label{lagr}
\eeq
Since the chemical potential is nonzero, 
${\tilde A}_0$ has an  expectation value, 
and this term plays the role of a tachyonic mass for the scalars
(see, e.g., \cite {Linde}). When  $\langle  2e {\tilde A}_0 \rangle = m_H$, 
the scalar field condenses. We look at a solution with no boundary charge,
for which $\langle  {A}_0 \rangle = 0$,  $\mu_s=m_H$ and  
\beq
\langle \sigma \rangle \equiv \sigma_c  = \sqrt{\frac{J_0}{2 m_H}} \,.
\label{b0prime}
\eeq
Hence, we're looking at  the  effects  that are $1/ m_H$  suppressed. 

\vspace{0.3cm}
\begin{center}
{\large \bf 3. Static Potential Without Fermion Fluctuations}
\end{center}
\vspace{0.3cm}

The uniform background solution obtained above sets a preferred Lorentz frame.
Let us look at small perturbations in this frame. 
We introduce  perturbations of the gauge, $A_\mu \equiv b_\mu$,  
and the scalar field,  as follows:
\beq
\sigma(x) = 
\sigma_c + \tau(x)\,.
\label{pert}
\eeq
Let us for the time being ignore fermion fluctuations. Physically this 
would correspond to the case when fermions are ``frozen in'' 
(for instance if they form a rigid-enough crystalline structure). 
We will relax this conditions and  include their fluctuation 
in the next Section.

Then, the Lagrangian density for the fluctuations 
in the quadratic approximation reads \cite {GGRR1}
\beq
{\cal L}_{2}= -{1\over 4} f_{\mu\nu}^2 + {1\over 2} (\partial_{\mu} \tau)^2+
{1\over 2} m_\gamma^2 b_\mu^2  + 2m_Hm_\gamma \,b_0 \tau \,.
\label{2lagr}
\eeq
Here 
\beq
m_\gamma^2 \equiv (2e)^2 {J_0\over 2m_H}\,,
\label{photonmass}
\eeq
$f_{\mu\nu}$ denotes the field strength for $b_\mu$, and 
we left out  for simplicity all the fermionic terms.  
It is useful to integrate out the $\tau$ field from 
(\ref {2lagr}). The remaining Lagrangian  takes the form:
\beq
{\cal L}_{2}= -{1\over 4} f_{\mu\nu}^2 +
{1\over 2} m_\gamma^2 b_\mu^2 +
{1\over 2} \,b_0 
{(2m_Hm_\gamma)^2\over \square}b_0. 
\label{Ltau}
\eeq
This Lagrangian contains four components of $b_\mu$, 
and no other fields. The first two terms in 
(\ref {Ltau}) are those of a usual massive photon with three degrees 
of freedom. The last term is unusual, as it gives rise to the 
dynamics to the timelike component of the gauge field. This term 
emerged due to the mixing of $b_0$ with the dynamical field 
$\tau$ in (\ref {2lagr}), and since we integrated out $\tau$, 
$b_0$ inherited its dynamics in a seemingly nonlocal way. 
The fact that there are no pathologies in 
(\ref {2lagr}) or (\ref {Ltau}), such as ghost and/or tachyons,  
can be seen by calculating the Hamiltonian density from (\ref {2lagr}):
\beq
{\cal H}= {\pi_j^2  \over 2} 
+ {f^2_{ij}\over 4}  + {(\partial_j \pi_j - 2 m_H m_\gamma \tau)^2\over 
2m_\gamma^2 }+
{P_\tau^2+(\partial_j\tau)^2 \over 2}\,.
\label{ham}
\eeq
Here,  $\pi_j \equiv -f_{0j}$ and $P_\tau\equiv \partial_0\tau$, and 
the Hamiltonian is positive semi-definite.

\vspace{0.3cm}

The form of the Lagrangian (\ref {Ltau}) is most useful for 
calculating a propagator. Indeed, the inverse of the quadratic 
operator that appears in 
(\ref {Ltau}) has poles which describe all the four propagating 
degrees of freedom. This propagator, sandwiched between two conserved 
currents $J_\mu$ and  $J^\prime_\mu$,  takes the form:
\beq
J_0 \left (  
-p^2 +m_\gamma^2 + \left [1- {\omega^2\over m_\gamma^2} \right ] 
{4M^4\over -p^2} \right )^{-1} a J_0^\prime 
- J_j \left (  -p^2 +m_\gamma^2 \right )^{-1}  J_j^\prime \,.
\label{amplitude}
\eeq
This describes two transverse photons with the mass $m_\gamma$,  
one heavy mode with mass $2m_H$, and a light phonon (for their dispersion 
relations, without and with fermion fluctuations, see   
Refs. \cite {GGRR1,GGRR3} respectively). Here 
$a\equiv 1- 4M^4\omega^2/p^2 m_\gamma^2(p^2-m_\gamma^2)$.

In particular, we are interested in a static potential, which takes the 
form
\beq
G(\omega=0, \k) = \left ( \k^2 +m_\gamma^2 + {4M^4\over \k^2} \right 
)^{-1}\,,
\label{static}
\eeq
where, we have used the notation $M\equiv\sqrt{m_H m_\gamma}$, 
and the first, second, and third terms on the r.h.s. of 
(\ref {static}) are due to the respective terms in (\ref {Ltau}).

Interestingly, when $m_H \gg m_\gamma $, as is the case here,  
there is no scale at which the photon mass 
term in  (\ref {static}) would dominate:   for $\k^2 \gsim m^2_\gamma$ 
the mass term is sub-dominant to the $\k^2$ term, while  
for $\k^2 \lsim m^2_\gamma $  it is sub-dominant to 
the $M^4/\k^2$ term.  

Let us  rewrite  the Green's  function (\ref {static}) in the following form:
\beq
G(\k, \omega=0)= \left ( \k^2 +m_\gamma^2 \right )^{-1} - \left ( 
\k^2 +m_\gamma^2+ {\k^2 (\k^2+m_\gamma^2)^2\over 4M^4} \right )^{-1}\,.
\label{photonphonon}
\eeq
The first term on the r.h.s. of (\ref {photonphonon})  could be 
thought of as an instantaneous 
repulsive screened Coulomb (Yukawa) potential, while  the 
second term as an attractive  potential due to the phonon. 
The phonon here is somewhat peculiar -- it mixes with the 
timelike component of the gauge field, and thus acquires an 
instantaneous piece, a part of which cancels  the instantaneous 
gauge potential. This is reflected in (\ref {photonphonon}) as 
exact cancellation of the residues of the pole at $\k^2+m_0^2=0$.
The remaining instantaneous part is what's given in (\ref {static}).

As we have already mentioned in Section 1, the 
remaining two poles of the second term in  (\ref {photonphonon}), 
describe  respectively the heavy state of mass $2m_H$, 
which is unimportant for the low-energy dynamics, and a light state 
-- the phonon \cite {GGRR3}. Just to reiterate, for simplicity of 
discussions,  in this  work we're using a somewhat imprecise 
language  and calling  the whole second term 
in  (\ref {photonphonon}) the phonon contribution. 

The key observation is that at scales larger than  $1/M$, 
which are of the primary interest, the phonon potential cancels  
the gauge potential  with a high accuracy.
Note that this cancellation is reliable at scales that are 
much greater than $J_0^{-1/3}\gg M^{-1}$, and takes place already at 
scales that   are much shorter that the photon 
Compton wavelength $m^{-1}_\gamma \gg J_0^{-1/3}$!

In a Lorentz-invariant theory having a negative sign in front of a 
propagator, such as the one in the second term in (\ref {photonphonon}),
would suggest the presence of a ghost-like state. However, 
this is not the case in a Lorentz-violating theory 
described by our Lagrangian (\ref {2lagr}) or (\ref {Ltau}).
As we pointed out above, the Hamiltonian density of this theory is 
positive semi-definite, and hence, no ghost or tachyons are present.
Moreover,  consistently with ones expectation, 
the second term in (\ref {photonphonon}) disappears 
in the limit $M\to 0$,  where Lorentz invariance of 
(\ref {2lagr}) or (\ref {Ltau}) is restored.

The above described properties  can also be seen by calculating  
the coordinate space potential (see, \cite {GGRR3}): 
\beq
V \equiv (Q_1e Q_2e) 
\int {d^3 \k \over (2\pi)^3} e^{i\k {\bf x}}G(\omega=0, \k) 
\propto  {Q_1Q_2 \alpha_{\rm em} e^{-Mr}\over \,r}
{\rm cos} (Mr)\,,
\label{potential}
\eeq
in which we assumed that  $r=|{\bf x}|\gg 
1/M$ and $m_\gamma  \ll M$.  Such a potential  was first 
derived in \cite {GGRR3}. It is  sign-indefinite  and 
undergoes modulated oscillations between  repulsion and attraction.  
There are an infinite number of points in the position space where    
the force between classical  charges would vanish. 
These are points where  
\beq
{dV\over dr}(r=r_n)=0, ~~~~~~n=1,2,...
\label{po}  
\eeq
The potential wells described by (\ref {po}) are too shallow 
to introduce a significant quantum-dynamics for static impurities -- 
any two charged probe particles (e.g., nuclear hydrogen or helium-3 
impurities) separated by a distance $r_n \gsim J_0^{-1/3}$, where our 
calculations are reliable,   would stay  in a static 
equilibrium  as long as $V(r_n)<0$. This is because their charges
will be screened efficiently  by the ambient condensate.

\vspace{0.3cm}

Before turning  to the next section we make three important comments:

{\it Comment 1}: Concerns the decoupling  of the helium-4 sates in the 
$m_H\to \infty$ limit. As was already mentioned in Sections 1 and 2,
the decoupling takes place via ``freezing'' out
of the phonon associated with the collective motion of the heavy modes.
The fact that this indeed is the case can be seen by looking at 
the dispersion relation for the phonon which is obtained from the 
poles of the Green's function (\ref {static}) in which  
nonzero $\omega$ is restored. In  Refs. \cite {GGRR1} and \cite {GGRR3} 
the full dispersion relations for the phonon without and 
with fermion fluctuations was  given respectively. 
The ``freezing out''  of the phonon takes place in both 
cases. 

For simplicity we discuss here the case of \cite {GGRR3}:  
For the relevant momenta $M^2 \gg \k^2$ the dispersion 
relation  reduces to  $\omega \simeq  m_\gamma 
(1 +\k^2 (\k^2 -m_\gamma^2)/8M^4)$,  and the phonon group velocity
\beq
\label{vgr}
v_{\rm gr} \simeq  {m_\gamma |\k| (2\k^2 -m_\gamma^2)\over 4M^4}\,,
\eeq
vanishes in the $m_H \to \infty$ limit. 

Note that for  $\k^2 \simeq m_\gamma^2/2$ the phonon 
group velocity vanishes for finite $m_H$. This  describes a state of a 
nonzero momentum but zero group velocity. The energy of this state 
is also nonzero, and to a good approximation equals to $m_\gamma$.
These properties are  similar to those of a roton
in superfluid helium II. Moreover, for excitations with $\k^2 > m_\gamma^2/2$
the group velocity is positive, while in the opposite case, 
$\k^2 < m_\gamma^2/2$,  it becomes  negative (i.e., the direction of 
the momentum and that of group velocity are opposite to each other).  
These excitations resemble the positive and negative group velocity 
rotons in superfluid helium II.

\vspace{0.2cm}

{\it Comment 2}: Concerns the issue of applicability of the 
linearized approximation (\ref {2lagr}) that we adopted in this work. 
For this we should restore back all the non-linear 
terms  that were ignored in (\ref {2lagr}) and compare them with
the linear ones retained there. Although this can 
be done in  full generality,  it is  instructive to look at the 
dynamics of a longitudinal polarization of the massive vector field
(i.e., the would-be  Nambu-Goldstone boson) which determines the fastest 
growing  non-linear terms in the Lagrangian at high energies 
\cite {Vainshtein,Levin}. This is easily achieved by employing the 
substitution $b_\mu \to (\partial_\mu n)/m_\gamma $, and taking the 
decoupling limit:   $m_\gamma \to 0~{\rm and}~e\to 0$,  while keeping 
the ratio $v\equiv m_\gamma/e  $ finite and fixed.
The resulting Lagrangian for the longitudinal mode $n$,  
and the remaining field $\tau$ takes the form
\beq
{ (\partial_\mu n)^2   \over 2} + {(\partial_\mu n)^2 \tau \over v}+
{(\partial_\mu n)^2 \tau^2 \over 2 v^2}+
2 m_H (\partial_0 n) \tau + {m_H (\partial_0 n) \tau^2 \over v}+
{(\partial_\mu \tau)^2   \over 2}\,.
\label{NG}
\eeq 
From this we deduce that the non-linear terms become 
comparable with the linear ones when $\tau \sim v$; we also  
require that fluctuations of gauge field to be 
smaller tat the  scalar chemical potential, which 
gives  $\partial_0 n\sim m_Hv$. Thus, the domain 
of applicability  of the linearized results of this work is 
\beq
\tau\ll v= \left ({J_0 \over 2 m_H}  \right )^{1/2}, ~~~~
\partial_0 n \ll  m_Hv = \left ({J_0  m_H}  \right )^{1/2}\,.
\label{limits}
\eeq
In the limit $m_H\to \infty$, 
the domain of applicability of the linearized  results  shrinks to zero.
This suggests  that the geometric size of the region in which 
one can meaningfully talk about  the charged condensate  
should be greater than a certain critical size that scales as 
$\sim v^{-1}$. The latter tends to infinity as $m_H\to \infty$.

\vspace{0.2cm}

{\it Comment 3:}  The dynamics of mall fluctuations 
can also be obtained from the effective Lagrangian 
(\ref {Leff1}). For this we restrict ourselves to the quadratic order 
in the expansion of ${\cal P} (x)$
\beq
{\cal L}_{eff} =  2e \Sigma^2 B_0 - 
{( \nabla_j  \Sigma )^2 + (2e)^2 \Sigma^2 B_j^2  \over 2m_H} 
+ C_2 (2e)^2 \Sigma^4 B^2_0\,,
\label{Leff2}
\eeq
and expand the above expression (amended with the gauge-kinetic and 
fermionic terms) to the second order in the following fluctuations:
\beq
\Sigma =  \sqrt{J_0\over 2} + \sqrt{m_H} \tau(x)\,,~~~~~   
B_\mu \equiv  b_\mu\,.
\label{flucSigma}
\eeq
The resulting Lagrangian for the fluctuations reads:
\beq
{\cal L}_{2}= -{1\over 4} f_{\mu\nu}^2  - {1\over 2} (\partial_{j} \tau)^2+
{1\over 2} m_0^2 b_0^2 - {1\over 2} m_\gamma^2 b_j^2 + 2m_H m_\gamma 
\,b_0 \tau \,,
\label{3lagr}
\eeq
where $m_0^2 = 2e^2 C_2J_0^2$ and, as before, 
$m_\gamma^2 = (2e)^2J_0/2m_H$.  
This should be compared with 
(\ref {2lagr}). We notice that unlike in  (\ref {2lagr}), 
there is no kinetic  term for $\tau$ in (\ref {3lagr}). 
This is because we decoupled the 
time-dependence of the heavy field in (\ref {Leff})
in the non-relativistic limit; thus, 
the $\Sigma$ field,  or its fluctuation $\tau$, are left to be 
instantaneous. This, obviously, does not change 
the conclusions  on  the  static potential discussed in 
this Section.

Furthermore, depending on the coefficient $C_2$ the ``electric'' 
and ``magnetic'' masses of the gauge field can be different.   
The value of $C_2$ cannot be fixed within  the effective 
Lagrangian approach. One can only impose a bound on it. For instance, 
without the $b_0-\tau$ mixing, superluminal group velocity  
is avoided when \footnote{Note that the longitudinal part of $b_j$ is 
a propagating mode. Imposing transversality on  $b_j$ would 
contradict general equations of motion. Also, superluminal group velocity 
in the present case would suggest some instability of the background.
In general,  superluminal group velocities can exist in 
highly absorptive media, in which case group velocity does not 
approximate the velocity of  propagation of a signal.}
\beq
m_0^2 \ge m_\gamma^2\,.
\label{Lfluct}
\eeq
In this Section  the value of $C_2$ was assumed to saturate  
the above inequality, as we put $m_0^2=m^2_\gamma $. 
However,  the fermion fluctuations,  which we ignored so far,
will  produce a large  hierarchy between the coefficients of 
the $b_0^2$ and  $b_j^2$ terms
in the Lagrangian, even if these were assumed to be equal in the 
classical theory.  

\vspace{0.1cm}

Finally, let us look at the decoupling limit: 
$e\to 0$,   while $m_0/e$ and  $m_\gamma/e$  are finite and   
fixed. Then, the Lagrangian (\ref {3lagr}) reduces to
\beq
{ m_0^2  \over 2m_\gamma^2} 
(\partial_0 n)^2  -   {1\over 2} (\partial_j n)^2 +   
2 m_H (\partial_0 n) \tau  -  
{1  \over 2}(\partial_j \tau)^2\,.
\label{NGeff}
\eeq 
This should be compared with the quadratic part of 
(\ref {NGeff}). Again,  there is missing kinetic term for 
$\tau$, for reasons explained above. 
As before, $\tau$ can be integrated out.  
The resulting dispersion relation for the remaining  mode reads
\beq
\omega^2 \simeq {\k^4 \over 4m_H^2 + \k^2 (m_0^2/m_\gamma^2)}\,,
\label{ok2}
\eeq
which for small momenta, $\k^2 \ll m_H^2(m_\gamma^2/m_0^2)$, 
gives $\omega \simeq \k^2/2m_H$, as it should be the case for 
a collection of free massive particles.

\vspace{0.3cm}
\begin{center}
{\large \bf 4. Quantum Screening of Static Charges }
\end{center}
\vspace{0.3cm}

In the previous section the fermions were treated as ``frozen''.
In may physical circumstances, and in particular in white dwarfs,  
this is not a good approximation. For the helium-electron system 
the fermion  fluctuations  should be taken into account. 
This was done in Ref. \cite {GGRR3} 
using the Thomas-Fermi (TF) approximation. The result 
is that the timelike component of the gauge field 
acquired an additional mass -- the electric mass. 
This breaks the degeneracy between the coefficients of 
$b_0^2$ and $b_j^2$ discussed in the previous Section.

However, the TF approximation  does not capture one  
significant property of the fermion system  related 
to a possibility of  exciting  gap-less modes near the Fermi surface,
which we will incorporate below  into  our results by using the  
one-loop correction to the propagator of 
(\ref {Ltau}). For this, we  restore back in the Lagrangian 
(\ref {Ltau}) the fermion  kinetic, mass and chemical potential
terms and, upon calculating the gauge boson propagator, 
will take into account the known one-loop gauge boson 
polarization diagram.

Since we're interested in a static potential, we look at 
the $\{00\}$ component of the propagator $ D_{00}$, and the static 
potential obtained from it:  
\beq
\tilde V(k)\equiv - D_{00}(\omega =0, \k)=
\left ( \k^2 +m_\gamma^2 +  {4M^4\over \k^2} 
+ F (k^2,k_F,m_f)  \right )^{-1}\,.
\label{V00}
\eeq
where  the function $F (k^2,k_F,m_f)$, which  is due to the one-loop 
photon polarization diagram, includes both the vacuum and 
fermion matter contributions ($k_F$ denotes the Fermi momentum). 
A complete expression for 
$F (k^2,k_F,m_f)$ can be found in Ref. \cite {KapustaT}. 
We concentrate   on  the expression for $F (k^2,k_F,m_f)$ in 
the massless ($m_f=0$) limit that is a good approximation for 
ultra-relativistic fermions:
\beq
F (\k^2,k_F)=  {e^2 \over 24\pi^2}  
\left ( 
16k^2_F +{k_F (4k_F^2 -3k^2)\over k}\ln ( {2 k_F  +\k \over 2 k_F -\k })^2
- k^2 \ln ( { k^2 -4 k_F^2\over \mu_0^2} )^2 
\right )\,.  
\label{F}
\eeq
Here $\mu_0$  stands for the normalization point that appears in the 
one-loop vacuum polarization diagram calculation. The  function $F$ 
introduces a  shift  of the pole in the propagator, corresponding to the
``electric mass'' of the photon. This part of the pole can be incorporated  
via the TF approximation, as it was done  in 
\cite {GGRR3}.  In addition, however,  the function $F$ 
also  gives  rise to branch cuts in the complex $|\k|$ 
plane (see \cite {Walecka} for the list of earlier references 
on this).
 
As in the previous section we can decompose  the static potential 
as follows: 
\beq
\tilde V (\k, \omega=0)= \left ( \k^2 +m_\gamma^2 +F\right )^{-1} - \left ( 
\k^2 +m_\gamma^2+ F+ {\k^2 (\k^2+m_\gamma^2+F)^2\over 4M^4} \right )^{-1}\,.
\label{photonphonon1}
\eeq
Note that the first term in (\ref {photonphonon1}) is 
just the instantaneous screened-Coulomb (Yukawa) potential  
of a massive photon with the one-loop 
polarization correction.  Our main interest is at 
distances smaller  than $m_\gamma^{-1}$. A sphere of radius 
$m_\gamma^{-1}$ encloses  many particles within its  volume
since $m_\gamma^{-1}\gg J_0^{-1/3}$.  At these scales, the 
first term in (\ref {photonphonon1}) can be approximated by:
\beq
{1\over \k^2 +F}\,.
\label{CoulF} 
\eeq
The above expression has a regular pole corresponding to the  acquired 
``electric'' mass of the photon  due to the polarization diagram. 
The contribution of this  pole would give rise to an exponentially 
decaying potential $e^{-m_{\rm el}r}/r$, where $m_{\rm el} \sim e \mu_f$. 
This is just an ordinary Debye screening.

However, as was mentioned  above,  
the expression (\ref {CoulF}) also has
branch cuts in the complex $|\k|$ plane for $k=\pm 2k_F$. These branch cuts 
give rise to the additional terms in  the static potential which are not 
exponentially suppressed,  but instead have an oscillatory behavior 
with a   power-like decaying envelope. In  a 
non-relativistic theory they're known as the Friedel 
oscillations \cite {Walecka}. In the relativistic theory  they were 
calculated  in Refs. \cite {Sivak,KapustaT} (we follow here 
\cite {KapustaT} and for simplicity ignore the running of the 
coupling constant due to the vacuum loop):
\beq
\Delta V= {Q_1Q_2 \alpha_{\rm em}^2 
\over 4\pi} {{\rm sin}(2k_Fr)\over k_F^3r^4}\,.
\label{oscp}
\eeq
Note that these branch cuts have a physical interpretation:
Since there is no mass gap in the fermion spectrum, 
a photon  can produce a near-the-Fermi-surface 
particle-hole  pair of an arbitrarily  small energy and the momentum 
close to $\pm2k_F$. The imaginary part of the one-loop photon 
polarization diagram  should include the continuum of such 
near-the-Fermi-surface pairs. These are reflected as logarithmic 
branch cuts in the expression for $F$.

Thus, if the phonon term (the second term) on the r.h.s. of 
(\ref {photonphonon1}) were absent one would have a power-like 
behavior (\ref {oscp}) of the static potential at scales 
$r\lsim m_\gamma^{-1}$. The phonon term, however,  significantly reduces  
the strength of this potential. The result for it could be calculated 
by directly taking Fourier transform of (\ref {V00}).  
The dominant contribution comes from the branch cuts 
at $k=\pm 2k_F$. Drawing  the contours  around these cuts 
in the upper half plane of complex $|\k|$ \cite {Walecka,KapustaT},  
one deduces  the result. In the approximation  
$M\gg k_F \gg m_\gamma$, which is  relevant for our system, a 
static potential between like charges scales as 
\beq
\Delta V \simeq  {4 Q_1Q_2 \alpha_{\rm em}^2 
\over \pi} {k_F^5~ {\rm sin}(2k_Fr)\over M^8 r^4}\,.
\label{oscpo}
\eeq
This potential has an additional suppression  factor of 
$16(k_F/M)^8\sim 10^{-23}$, as compared to $\Delta V $   in  a theory 
without phonons (\ref {oscp}).   Nevertheless, 
$\Delta V $  in (\ref {oscpo})  dominates over the 
exponentially suppressed part of the total potential
found in Ref. \cite {GGRR3} and discussed in the previous 
section (see,  (\ref {potential})), for separations between 
probe particles large-enough for the effective field theory 
description to be  applicable. The net static interaction in the 
charged condensate, set by  (\ref {oscpo}),  
is very weak.  It is, however,  still much stronger 
than gravitational interaction between a pair of light nuclei.

Although formally  $\Delta V $  in (\ref {oscpo}) 
is proportional to $\alpha_{\rm em}^2$,  to a good approximation 
it is  independent of  $\alpha_{\rm em}$ 
since $M^8 \propto e^4 (m_H J_0)^2$.

As in the previous section, the  potential for the static charges 
(\ref {oscpo}) is  shallow and sign-indefinite.
There are, however,  an infinite number of position-points where    
the force between static charges would vanish.
These are determined by extremizing (\ref {oscpo})
and obtaining
\beq
4{\rm tan}(2k_Fr_n)= 2k_Fr_n, ~~~~n =1,2,3,...
\label{extr}
\eeq
Some of the $r_n$'s are local minima, and thus, static  charges (e.g.,
helium-4 nuclei, nuclear hydrogen or helium-3 impurities)  placed in 
those minima with $r_n \gsim J_0^{-1/3}$, where our approximations are valid,  
would stay  in a local equilibrium  due to efficient screening 
by the condensate. Hence, the condensate  could tolerate some 
fraction of charged impurities before the latter could significantly 
affect the condensate itself.

Finite temperature corrections would modify these results
quantitatively,  however, the main point of efficient screening of 
the static charges in the condensate should remain  valid at 
temperatures well-below the condensation point.
For instance, in white dwarfs with temperature $10^6 - 10^7\,K$
we would expect the dominant  temperature-dependent 
corrections to  the potential to be proportional to  
$T/ J_0^{1/3} \sim (10^{-4}-10^{-3})\ll 1$,  which are  
negligible.

\vspace{0.1cm}
\begin{center}
{\bf Acknowledgments }
\end{center}
\vspace{0.1cm}

GG is grateful to  Juan Maldacena for 
stimulating questions,  to  Arkady Vainshtein for helpful discussions, and 
especially to Nima Arkani-Hamed for valuable  conversations and comments. 
He also thanks A. Dolgov for sending the draft of \cite {Dolgov} 
before its publication.
The work of GG was partially  supported by  the NSF and NASA 
grants (PHY-0758032, NNGG05GH34G). RAR was supported by the 
James Arthur Graduate Assistantship at NYU.



\begin{thebibliography}{99}


\bibitem{GGRR1} G.~Gabadadze and R.~A.~Rosen,
Phys Lett. {\bf B} 658 (2008), 266
  [arXiv:0706.2304 [hep-th]]. 


\bibitem{GGRR3} G.~Gabadadze and R.~A.~Rosen,
  JCAP {\bf 0810}, 030 (2008)
  [arXiv:0806.3692 [astro-ph]].



\bibitem{Walecka} A.L. Fetter, J.D. Walecka, 
{\it ``Quantum Theory of Many-Particle Systems''},
McGraw-Hill, 1971. 


\bibitem{Sivak} H.D. Sivak, Physica {129\bf A}, 408 (1985).

\bibitem{KapustaT} 
J.~I.~Kapusta and T.~Toimela,
  Phys.\ Rev.\  D {\bf 37}, 3731 (1988).


\bibitem{Shafroth} M.R. Schafroth, Phys. Rev. {\bf 100}, 463 (1955). 

\bibitem{Foldy} L.L. Foldy, Phys. Rev. {\bf 124}, 649 (1961).


\bibitem{71} J.P. Hansen, B. Jancovici, D. Schiff, Phys. Rev. Lett.
{\bf 29}, 991 (1972).


\bibitem{Linde} A.~D.~Linde,
  Phys.\ Rev.\  D {\bf 14}, 3345 (1976).

\bibitem{Kapusta} J.~I.~Kapusta,
Phys.\ Rev.\  D {\bf 24} (1981) 426.

\bibitem{Haber} H.~E.~Haber and H.~A.~Weldon,
  Phys.\ Rev.\  D {\bf 25}, 502 (1982).

\bibitem{Dolgov} A.D. Dolgov, A. Lepidi, G. Piccinelli,  
arXiv:0811.4406 (hep-th).

\bibitem{Polchinski} J.~Polchinski,
  arXiv:hep-th/9210046.

\bibitem{LL} E.M. Lifshitz, L.P. Pitaevskii, ``Statistical Physics'', 
part 2,  (1980).

\bibitem{GWW}  M.~Greiter, F.~Wilczek and E.~Witten,
  Mod.\ Phys.\ Lett.\  B {\bf 3}, 903 (1989).


\bibitem{Vainshtein} 
A.~I.~Vainshtein and I.~B.~Khriplovich,
  Yad.\ Fiz.\  {\bf 13} (1971) 198.

\bibitem{Levin} 
J.~M.~Cornwall, D.~N.~Levin and G.~Tiktopoulos,
  Phys.\ Rev.\  D {\bf 10}, 1145 (1974)
  [Erratum-ibid.\  D {\bf 11}, 972 (1975)].



\end{thebibliography}
\end{document}